\documentclass[]{aa}
\usepackage{psfig}
\begin{document}

\thesaurus{07(07.13.2)}

\title{CCD photometry and new models of 5 minor planets}

\author{L.L. Kiss\inst{1,3} \and Gy. Szab\'o\inst{1,3} \and
K. S\'arneczky\inst{2,3}}

\institute{Department of Experimental Physics \& Astronomical Observatory,
JATE University, H-6720 Szeged, D\'om t\'er 9., Hungary
\and Department of Physical Geography, ELTE University, H-1088 Budapest,
Ludovika t\'er 2., Hungary
\and
Guest Observer at Konkoly Observatory}

\titlerunning{CCD photometry and new models of 5 minor planets}
\authorrunning{Kiss et al.}
\offprints{l.kiss@physx.u-szeged.hu}
\date{}

\maketitle

\begin{abstract}

We present new $R$ filtered CCD observations of 5 faint and moderately
faint asteroids carried out between October, 1998 and January, 1999.
The achieved accuracy is between 0.01--0.03 mag, depending mainly
on the target brightness.
The obtained sinodic periods and amplitudes:
683~Lanzia -- $4\fh6\pm0\fh2$, 0.13 mag; 725~Amanda -- $>3\fh0$,
$\geq$0.40 mag;
852~Wladilena -- $4\fh62\pm0\fh01$, 0.32 mag (December, 1998) and 0.27 mag
(January, 1999); 1627~Ivar -- $4\fh80\pm0\fh01$,
0.77 mag (December, 1998) and 0.92 mag (January, 1999).
The Near Earth Object 1998~PG unambiguously showed doubly-periodic
lightcurve, suggesting the possibility of a relatively fast
precession (P$_1$=1\fh3, P$_2$=5\fh3).

Collecting all data from the literature, we determined new models
for 3 minor planets. The resulting spin vectors and
triaxial ellipsoids have been calculated by an amplitude-method.
Sidereal periods and senses of rotation were calculated
for two asteroids (683 and 1627) by a modified epoch-method.
The results are:
683 -- $\lambda_p=15/195\pm25^\circ$,
$\beta_p=52\pm15^\circ$, $a/b=1.15\pm0.05$, $b/c=1.05\pm0.05$,
P$_{sid}$=0\fd1964156$\pm$0\fd0000001, retrograde;
852 -- $\lambda_p=30/210\pm20^\circ$,
$\beta_p=30\pm10^\circ$, $a/b=2.3\pm0.3$, $b/c=1.2\pm0.2$;
1627 -- $\lambda_p=145/325\pm8^\circ$,
$\beta_p=34\pm6^\circ$, $a/b=2.0\pm0.1$, $b/c=1.09\pm0.05$,
P$_{sid}$=0\fd1999154$\pm$0\fd0000003, retrograde.
The obtained shape of 1627 is in good agreement with
radar images by Ostro et al. (1990).

\end{abstract}

\section{Introduction}

Ground-based modelling of the shape and the rotation of the minor planets
requires
high precision and long-term photometric observations. With the advent
of the CCD era it has become possible to study much fainter minor
planets than previously. The photometric methods of modelling
are based on multi-opposition lightcurves of full phase coverage
obtained at very different
longitudes (De Angelis 1993, Detal et al. 1994, Szab\'o et al. 1999).
Another important
aspect is to detect possible collisional effects, e.g. multiperiodic
lightcurves due to binarity or precession, as they may yield insights
into the recent solar system evolution.

We started a long-term observational project addressed to
photometric monitoring of selected minor planets. The observing
programme consists of asteroids with available multi-opposition lightcurves
enabling application of different photometric methods in order
to model their shape and rotation. First results of this project
have already been published in S\'arneczky et al. (1999) and
Szab\'o et al. (1999).
The main aim of this paper is to present new CCD observations
carried out between October, 1998 and January, 1999 and
models for 3 minor planets.
Observations, their
limitations and applied methods are discussed in Sect.\ 2,
while Sect.\ 3 deals with the detailed observational results.

\section{Observations and modelling methods}

We carried out $R_C$ filtered CCD observations at Piszk\'estet\H o
Station of Konkoly Observatory on ten nights from October, 1998
to January, 1999. The data were obtained using the 60/90/180~cm
Schmidt-telescope equipped with a Photometrics AT200 CCD
camera (1536x1024 KAF 1600 MCII coated CCD chip). The projected
sky area is 29'x18' which corresponds to an angular resolution
of 1\farcs1/pixel.

The exposure times were limited by two factors: firstly, the
asteroids were not allowed to move more than the
FWHM of the stellar profiles (varying from night to night) and
secondly, the signal-to-noise (SN) ratio had to be at least 10.
This latter parameter was estimated by comparing the peak pixel values
with the sky background during the observations.
The journal of observations is summarized in Table 1.

% Table 1.
\begin{table*}
\caption{The journal of observations. ($r$ -- geocentric distance;
$\Delta$ -- heliocentric distance; $\lambda$ -- ecliptic longitude;
$\beta$ -- ecliptic latitude; $\alpha$ -- solar phase angle;
aspect data are referred to 2000.0)}
\begin{center}
\begin{tabular} {llrrllrr}
\hline
Date & RA & Decl. & $r$(AU) & $\Delta$(AU) & $\lambda$ & $\beta$ & $\alpha$\\
\hline
{\bf 683 Lanzia} & & & & & & & \\
1998 12 14/15 & 00 12.78 & +19 58.9 & 3.25 & 2.82 & 20 & 18 & 17\\
1998 12 16/17 & 00 13.78 & +19 49.5 & 3.25 & 2.84 & 20 & 18 & 17\\
{\bf 725 Amanda} & & & & & & & \\
1999 01 26/27  & 06 31.61 & +27 13.5 & 2.34 & 1.44 & 110 & 24 & 12\\
{\bf 852 Wladilena} & & & & & & &\\
1998 12 12/13 & 11 40.37 & +27 28.2 & 2.98 & 2.71 & 163 & 23 & 19\\
1998 12 14/15 & 11 41.67 & +27 31.9 & 2.98 & 2.68 & 163 & 23 & 19\\
1998 12 16/17 & 11 42.89 & +27 36.2 & 2.98 & 2.65 & 163 & 23 & 19\\
1999 01 24/25 & 11 47.52 & +30 52.9 & 2.95 & 2.18 & 170 & 19 & 14\\
{\bf 1627 Ivar} & & & & & & &\\
1998 12 14/15 & 05 03.41 & +10 30.4 & 2.22 & 1.26 & 76 & $-$12 & 6\\
1998 12 15/16 & 05 02.00 & +10 32.6 & 2.23 & 1.26 & 76 & $-$12 & 6\\
1998 12 16/17 & 05 00.62 & +10 34.9 & 2.23 & 1.27 & 76 & $-$12 & 7\\
1999 01 22/23 & 04 30.95 & +13 09.2 & 2.35 & 1.65 & 85 & $-$13 & 20\\
{\bf 1998 PG} & & & & & & &\\
1998 10 23/24 & 23 47.69 & +09 15.0 & 1.23 & 0.26 & 2  & 9  & 25\\
1998 10 26/27 & 23 55.11 & +08 26.9 & 1.23 & 0.27 & 2  & 7  & 25\\
1998 10 27/28 & 23 57.63 & +08 11.6 & 1.23 & 0.27 & 2  & 7  & 26\\
\hline
\end{tabular}
\end{center}

\end{table*}

The image reduction was done with standard IRAF routines.
The relatively high electronic noises and low angular resolution
did not permit the use of psf-photometry and that is why a simple aperture
photometry was performed with the IRAF task {\it noao.digiphot.apphot.qphot}.
Unfortunately other filters were not available during the
observing run and consequently we could obtain only instrumental
differential $R$ magnitudes in respect to closely separated comparison
stars. The precision
was estimated with the rms scatter of the comp.$-$check magnitudes
(tipically 0.01--0.03 mag).

We have also investigated the possible colour effects in neglecting
standard photometric transformations. We made an
$R$ filtered 60-seconds
CCD image of open cluster M67 on December 14, 1998. This cluster contains
a widely used sequence of photometric standard stars (Schild 1983).
We determined the instrumental magnitude differences in respect to
star No. 81 in Schild (1983), which various colour indices are close
to zero ($(B-V)=-0.098$, $(V-R)_C=-0.047$ mag). The studied standards
were stars No. 106, 108, 117, 124, 127, 128, 129, 130, 134 and 135,
following Schild's notation.
We plotted the resulting differences ($\Delta R_{ins}-\Delta R_{std}$)
vs. $(B-V)$ and $(V-R)_C$ in Fig.\ 1. For a wide colour range they
do not differ more than 0.1 mag, while the colour dependence is quite
weak. Therefore, the obtained instrumental $R$-amplitudes of minor planet
lightcurves are very close
to the standard ones, allowing reliable comparison with other
measurements.

%Fig. 1.
\begin{figure}
\begin{center}
\leavevmode
\psfig{figure=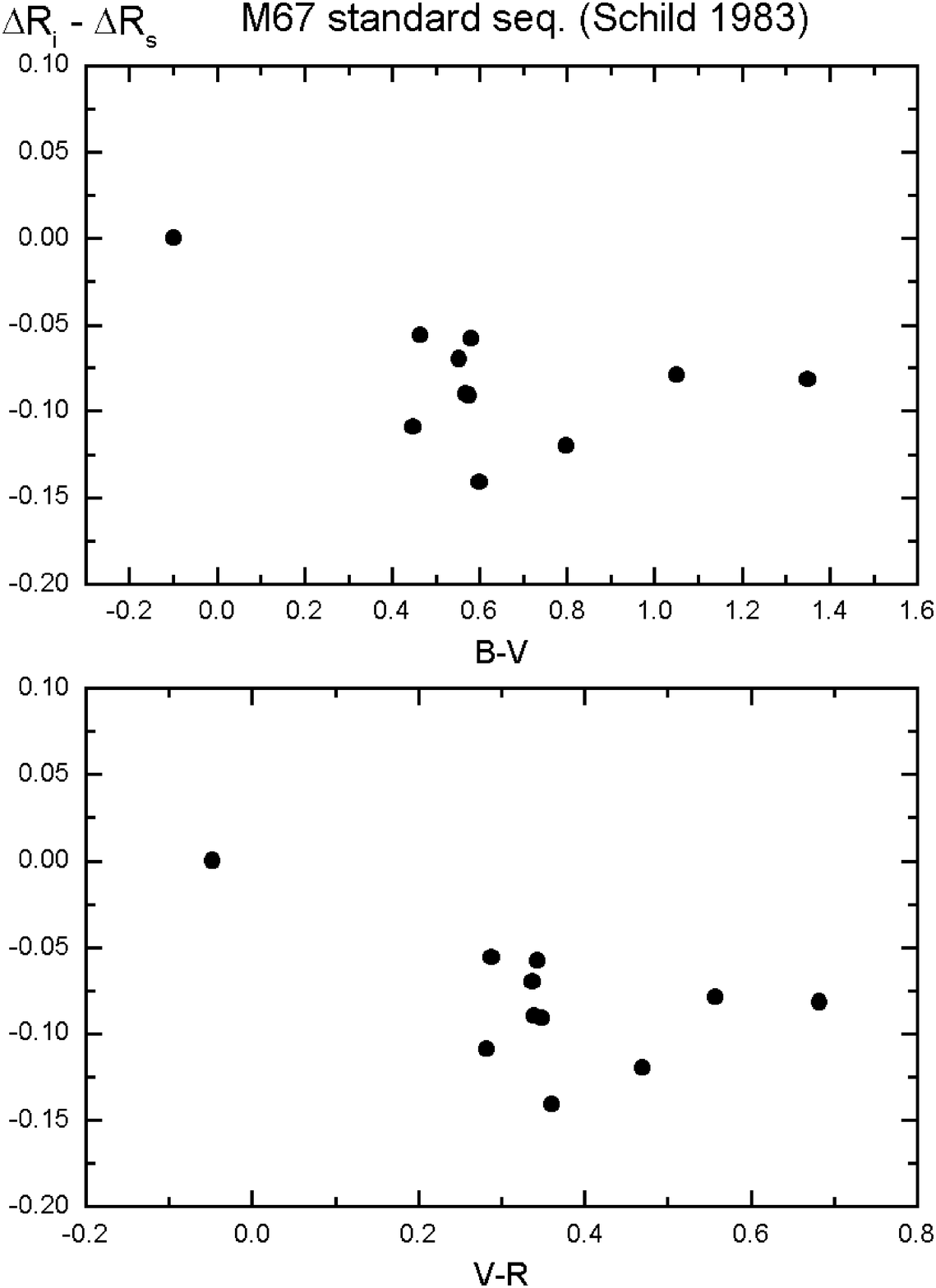,width=\linewidth}
\caption{The colour dependence of instrumental minus standard magnitude
differences for selected photometric standard stars in M67.}
\end{center}
\label{r_obs}
\end{figure}

The presented magnitudes throughout the paper are based on magnitudes
of the comparison stars taken from the Guide Star Catalogue (GSC) (Table 2).
Therefore, their absolute values are fairly uncertain
(at level of $\pm$0.2--0.3 mag). Fortunately
it does not affect the
other photometric parameters needed in the minor planet studies, such
as the amplitude, time of extrema, or photometric period.
The final step in the data reduction was the correction
for the light time\footnote{Individual data are available upon
request from the second author ({\tt szgy@neptun.physx.u-szeged.hu})}.
Composite diagrams were calculated using APC11 by Jokiel (1990) and are
also light time corrected. Times of zero phase are included in
the individual remarks.

% Table 2.
\begin{table}
\caption{The comparison stars. The typical uncertainty in the
magnitude values is as large as $\pm$0.2-0.3 mag.}
\begin{center}
\begin{tabular} {lll}
\hline
Date & Comp. & $m$(GSC)\\
\hline
{\bf 683 Lanzia} & & \\
1998 12 14 & GSC 1182 337 & 15.3\\% & GSC 1182 1096 & 15.0 \\
1998 12 16 & GSC 1182 85 & 14.4 \\% & GSC 1182 354 & 14.6 \\
{\bf 725 Amanda} & & \\
1999 01 26 & GSC 1887 1325 & 12.3 \\% & GSC 1887 1634 & 13.6 \\
{\bf 852 Wladilena} & & \\
1998 12 12 & GSC 1984 2286 & 12.7 \\% & GSC 1984 2202 & 14.3 \\
1998 12 14 & GSC 1984 2516 & 12.0 \\% & GSC 1984 2431 & 15.0 \\
1998 12 16 & GSC 1984 2496 & 13.8 \\% & GSC 1984 2180 & 14.2 \\
1999 01 24 & GSC 2524 1778 & 12.6 \\% & GSC 2524 2148 & 12.2 \\
{\bf 1627 Ivar} & & \\
1998 12 12 & GSC 702 759 & 12.6 \\% & GSC 702 690 & 13.5 \\
1998 12 14 & GSC 689 1331 & 12.8 \\% & GSC 689 1233 & 14.3 \\
1998 12 16 & GSC 689 2101 & 12.6 \\% & GSC 689 2009 & 14.1 \\
1999 01 22 & GSC 681 519 & 13.7 \\% & GSC 681 63 & 14.1 \\
{\bf 1998 PG} & & \\
1998 10 23 & GSC 1170 1119 & 14.2 \\% & GSC 1170 375 & 13.9 \\
1998 10 24 & GSC 1171 632 & 14.3 \\% & 1171 1368 & 15.0 \\
1998 10 26 & GSC 1171 1424 & 14.5 \\% & GSC 1171 1485 & 14.6 \\
\hline
\end{tabular}
\end{center}
\end{table}

Two methods were applied for modelling. The first is the
well-known amplitude-method described, e.g., by Magnusson (1989)
and Micha\l owski (1993). For this
the amplitude information is used to determine the spin vector and the
shape. An important point is that the observed $A(\alpha)$
amplitudes at solar
phase $\alpha$ should be reduced to zero phase ($A(0^\circ)$), if possible,
by a simple linear transformation
in form of $A(\alpha)=A(0^\circ)(1+m \alpha)$. $m$ is a parameter, which has
to be determined individually and that can be difficult, or even
impossible if there are insufficient observations (Zappala et al. 1990).

The other possibility is to examine the times of light extrema
(``epoch-methods'', ``E-methods'').
In this paper a modified version was
used, which gives the sense of the rotation unambiguously.
The pole coordinates can be also estimated independently.
Further details
can be found in Szab\'o et al. (1999) and Szab\'o et al. (in prep.),
here we give only a brief description.

The initial idea is that the prograde and retrograde rotation
can be distinguished by following the virtual shifts of
moments of light extrema (e.g. times of minima). From a
geocentric point of view, a full revolution around the Earth
causes one extra rotational cycle to be added (retrograde rotation)
or subtracted (prograde rotation)
to the observed number of rotational cycles during that period.
The virtual shifts increase or decrease monotonically and
their cumulative change is exactly one period over one
revolution. Therefore, plotting the observed minus calculated (O$-$C)
times of minima versus the geocentric longitude, we get a
monotone function ascending or descending by the value of the period.
The definition of the observed O$-$C is as
follows:
\begin{equation}\label{eq1}
O-C=T_{\rm min}-(E_{\rm 0}+N\cdot P_{\rm sid})\equiv
\langle {\Delta T \over P_{\rm sid}} \rangle P_{\rm sid}
\end{equation}
\noindent where $T_{\rm min}$ means the observed time of minimum, $E_{\rm 0}$
is the epoch, while the $N$ integer number denotes the cycles
(e.g. the number of rotation) of $P_{\rm sid}$ period
between the observed extremum and the epoch. ${\Delta T}$ means
the time interval between $T_{\rm min}$ and $E_{\rm 0}$, and
$\langle \rangle$ denotes fractional part.
The theoretical O$-$C curve depends on the pole coordinates:
\begin{equation}\label{eq2}
O-C':={O-C \over P_{\rm sid}} = {1 \over \pi} {\rm arctan}{{\rm tan}(\Lambda-\lambda_p)
\over {\rm sin}(\beta_p-B)}
\end{equation}
\noindent where $\Lambda$ and $B$ denote geocentric longitude and
latitude; ${\lambda {\rm p}}$ and ${\beta {\rm p}}$ are the pole coordinates.

The main difference between the classical E-methods and
this O$-$C' method is that time dependence is transformed
into the geocentric longitude domain.
Because of the system's basic symmetries, the O$-$C' diagrams
are calculated for a half revolution and with the half sidereal period.
The fitting procedure consists of altering $P_{sid}$ until the observed times
of minima do not give a
monotone O$-$C' diagram showing an increase or decrease of exactly 1.
Fitting a theoretical curve (Eq.\ 2) to the observed points,
the pole coordinates can be also estimated.

\section{Discussion}

\noindent {\it 683 Lanzia}

\noindent This minor planet was discovered by M. Wolf in Heidelberg,
on July 23, 1909. It was observed in the 1979, 1982, 1983-1984,
1987 oppositions (Carlsson \& Lagerkvist 1981, Weidenschilling et al.
1990). Carlsson \& Lagerkvist (1981) determined a rotation period of
4\fh322 and an amplitude of 0.14 mag. On the other hand,
Weidenschilling et al. (1990) measured a period of 4\fh37
with an amplitude of 0.12 mag.

Our observations in 1998 suggest a period of 4\fh6$\pm$0\fh2
with an amplitude of 0.13$\pm$0.01.
Composite diagrams calculated with previously published
periods between 4\fh3-4\fh4 have much larger scatter.
The light-time corrected composite diagram is presented in Fig.\ 2.
The zero phase is JD 2451162.3169.

Based on earlier data (see Table 3), a new model has been
determined with amplitude method.
The observed amplitudes vs. ecliptic longitudes with the
fit are plotted in Fig.\ 3.
The resulting triaxial ellipsoid has the
following parameters: a/b=1.15$\pm$0.07, b/c=1.05$\pm$0.05,
while the spin vector's coordinates are $\lambda_p=15/195\pm25^\circ$,
$\beta_p=52\pm15^\circ$, respectively.
We could not reduce the observed amplitudes
to zero solar phase, since the actual value of $m$ parameter
(e.g. Zappala et al. 1990)
could not be estimated by the data sequence or asteroid classification.
Also we have to note that a mixture of $V$ and $R$ amplitudes was used,
thus the model should be considered as an approximate one.
The O$-$C' model has also been determined (Fig.\ 4). For reducing the errors,
lightcurves obtained between October, 1983 and February, 1984, were
composed and one time of minimum was determined from
this composite lightcurve. The resulting sidereal period
is $P_{sid}$=0\fd1964156$\pm$0\fd0000001 with
retrograde rotation.

%Fig. 2.
\begin{figure}
\begin{center}
\leavevmode
\psfig{figure=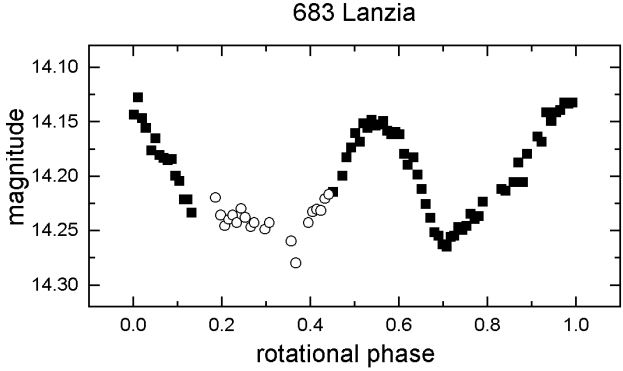,width=\linewidth}
\caption{The composite $R$ lightcurve of 683 (symbols: solid squares --
December 14; open circles -- December 16).}
\end{center}
\label{683}
\end{figure}

%Fig. 3.
\begin{figure}
\begin{center}
\leavevmode
\psfig{figure=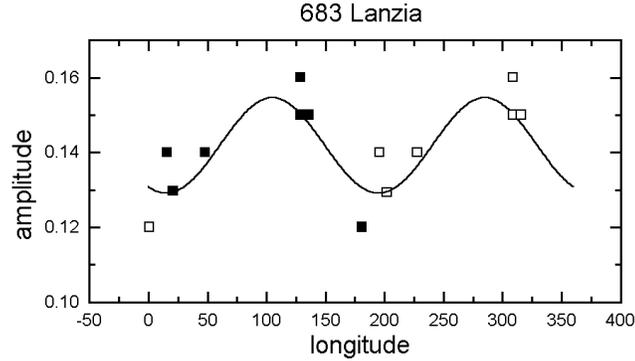,width=\linewidth}
\caption{The observed amplitudes vs. longitudes with the determined fit
for 683 Lanzia}
\end{center}
\label{683fit}
\end{figure}

%Fig. 4.
\begin{figure}
\begin{center}
\leavevmode
\psfig{figure=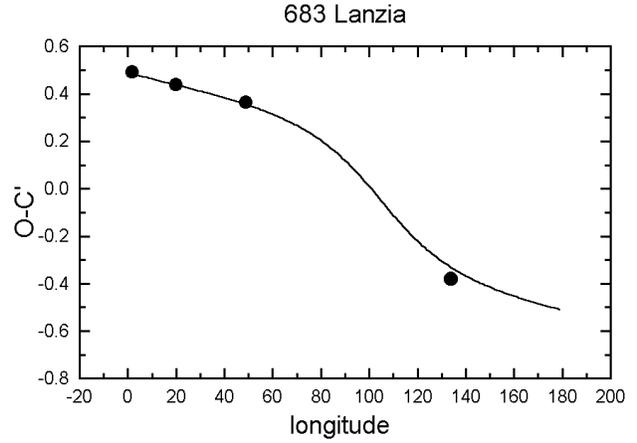,width=\linewidth}
\caption{The observed O$-$C' values fitted with the model for 683 Lanzia}
\end{center}
\label{683ocfit}
\end{figure}

%Table 3
\begin{table}
\begin{center}
\caption{Published photometries of 683 Lanzia}
\begin{tabular} {lrrrlll}
\hline
Date & $\lambda$ & $\beta$ & $\alpha$ & A & t$_{min}$ & ref. \\
\hline
1979 03 19,20  &     182   &   $-$27   &    9    &   0\fm12  & 43963.452 & (1)\\
1982 12 16     &     49    &    9    &    11   &   0.14    & 45319.591 & (2)\\
1983 10 12,13  &     130   &   $-$9    &    18   &   0.15  & 45650.871 & (2)\\
1983 11 15     &     137   &   $-$13   &    19   &   0.15  & 45650.871 & (2)\\
1984 02 21     &     129   &   $-$23   &    10   &   0.16  & 45650.871 & (2)\\
1987 10 19     &     16    &   23    &    7    &   0.12    & 47118.538 & (2)\\
1998 12 14,16  &     20    &   18    &    17   &   0.13   & 51162.275 & p.p.\\
\hline
\end{tabular}
\end{center}
References: (1) -- Carlsson \& Lagerkvist 1981; (2) -- Weidenschilling et al.
1990
\end{table}

\bigskip

\bigskip

\bigskip

\bigskip

\bigskip

\bigskip

\noindent {\it 725 Amanda}

\noindent It was discovered by J. Palisa in Vienna, on October 21, 1911.
To our knowledge, the only one photometry of 725 in the literature
is that of Di Martino et al. (1994) carried out in 1985. They determined
a sinodic period of 3\fh749 associated with a full variation of
0.3 mag. Our observations do not exclude that period, as they suggest
a possible value around 4 hours. Unfortunately the data cover only
3 hours, thus we could not draw a firm conclusion.
The observations were made under fairly unfavourable conditions,
which is illustrated with the comp--check curve bearing a relatively
high scatter (about $\pm$0.03 mag). It is presented together with
the observed lightcurve in Fig.\ 5.

%Fig. 5.
\begin{figure}
\begin{center}
\leavevmode
\psfig{figure=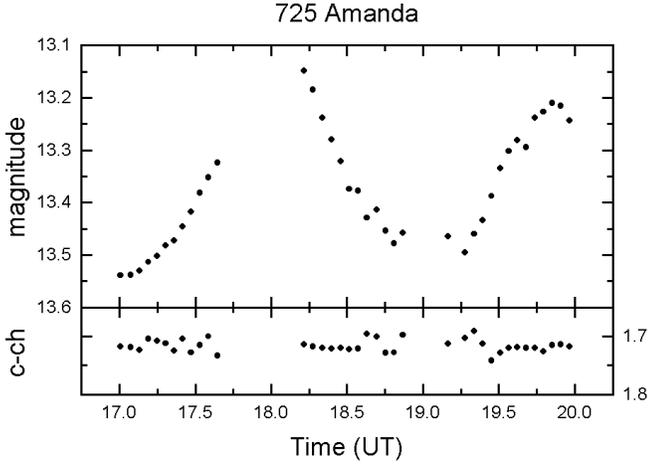,width=\linewidth}
\caption{The R lightcurve of 725 on January 26, 1999.}
\end{center}
\label{683}
\end{figure}

\bigskip
\noindent {\it 852 Wladilena}

\noindent This asteroid was discovered by S. Belyavskij in Simeis, on April 2,
1916. Its earlier photometric observations were carried out in 1977,
1982 and 1993 (Tedesco 1979, Di Martino \& Cacciatori 1984, De Angelis
\& Mottola 1995).
The observed light variation in 1998 had an amplitude of 0.32 mag,
while the period was 4\fh62$\pm$0\fh01. This is in very
good agreement with results by De Angelis \& Mottola (1995), who
found a period value of 4\fh613.
The light time corrected composite diagram is presented in Fig.\ 6. The
zero phase is at 2451160.5904.
The lightcurve has remarkable asymmetries -- the
brighter maximum is rather sharp, its hump is exactly two
times shorter than the other one. There are also
small amplitude, short-period humps on the longer descending branch.
These phenomena can be more or less identified in the previous
measurements too.
That is why we carried out a second observing run on January 24, 1999.
We wanted to check the reality of these irregularities.
The lightcurve revealed the same asymmetries as those
of observed one month earlier (Fig.\ 7). This may refer to a shape
with sharp asymmetries, e.g. something similar to a jagged tenpin.

%Fig. 6.
\begin{figure}
\begin{center}
\leavevmode
\psfig{figure=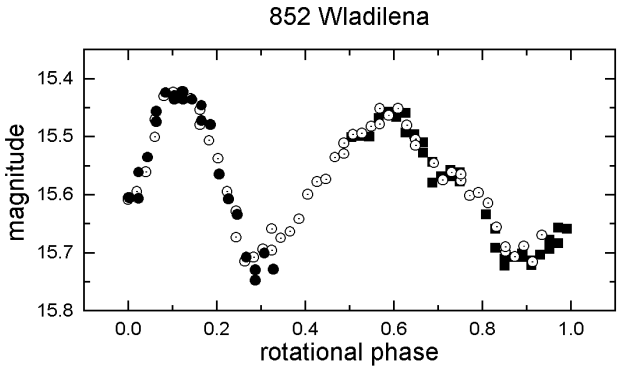,width=\linewidth}
\caption{The composite R lightcurve of 852 (symbols: solid circles --
December 12; dotted circles -- December 14; solid squares -- December 16).}
\end{center}
\label{852}
\end{figure}

%Fig. 7.
\begin{figure}
\begin{center}
\leavevmode
\psfig{figure=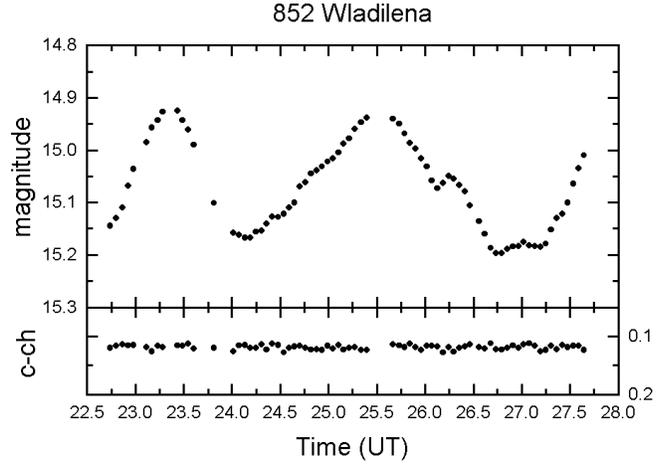,width=\linewidth}
\caption{The R lightcurve of 852 on January 24, 1999}
\end{center}
\label{852_99}
\end{figure}
We have tried to determine a new model using the earlier data
summarized in Table 4. Unfortunately, the measurements have
such a distribution along the longitude that
reliable modelling is difficult. This is shown in Fig.\ 8, where the observed
amplitudes
vs. ecliptic longitudes are plotted with an approximate fit.
The resulting parameters are as follows: a/b=2.3$\pm$0.3, b/c=1.2$\pm$0.2,
$\lambda_p=30/210\pm20^\circ$, $\beta_p=30\pm10^\circ$.
The pole coordinates are in considerable agreement with those
of by De Angelis \& Mottola (1995), who determined two possible
solutions: (1) $\lambda_p=53\pm6^\circ$, $\beta_p=24\pm20^\circ$
and (2) $\lambda_p=235\pm6^\circ$, $\beta_p=21\pm20^\circ$.

%Table 4
\begin{table}
\begin{center}
\caption{Published photometries of 852 Wladilena}
\begin{tabular} {lrrrll}
\hline
Date & $\lambda$ & $\beta$ & $\alpha$ & A & ref. \\
\hline
1977 02 14           &   139     &   31    &     10  &   1\fm12    & (1) \\
1982 10 18           &   6       &   $-$10   &     10  &   0.37    & (2) \\
1993 11 8,10         &   33      &   $-$8    &     3   &   0.23    & (3) \\
1998 12 12-16     &   163     &   23    &     19  &      0.32    & p.p. \\
1999 01 24      &   170     &   19    &     14  &   0.27    & p.p. \\
\hline
\end{tabular}
\end{center}
References: (1) -- Tedesco 1979; (2) -- Di Martino \& Cacciatori 1984;
(3) De Angelis \& Mottola 1995
\end{table}

%Fig. 8.
\begin{figure}
\begin{center}
\leavevmode
\psfig{figure=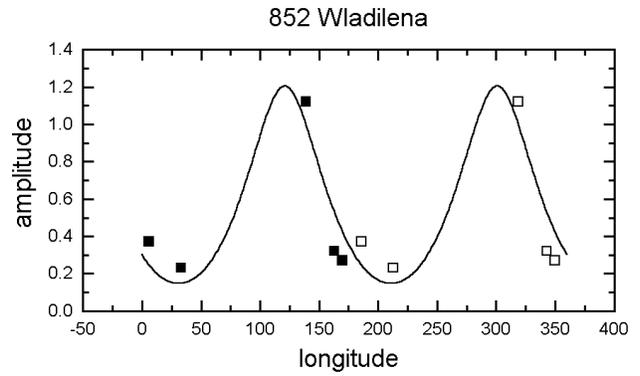,width=\linewidth}
\caption{The observed amplitudes vs. longitudes with
the determined fit for 852 Wladilena}
\end{center}
\label{852}
\end{figure}

\bigskip

\bigskip

\bigskip

\noindent {\it 1627 Ivar}

\noindent This Earth-approaching asteroid was discovered by
E. Hertzsprung in Johannesburg, on September 25, 1929.
There are four photometric observations
in the literature (Hahn et al. 1989,
Velichko et al. 1990, Hoffmann \& Geyer 1990, Chernova et al. 1995)
and one radar measurement by Ostro et al. (1990).
The previously determined periods scatter around 4\fh8,
thus our resulting
4\fh80$\pm$0.01 is in perfect agreement with earlier results.
The amplitude changed significantly over a period of one
month, as it was 0.77 mag and 0.92 mag in December, 1998 and
January, 1999, respectively.
The composite lightcurve is presented in Fig.\ 9, while
the single lightcurve obtained in January is plotted in Fig.\ 10.

%Fig. 9.
\begin{figure}
\begin{center}
\leavevmode
\psfig{figure=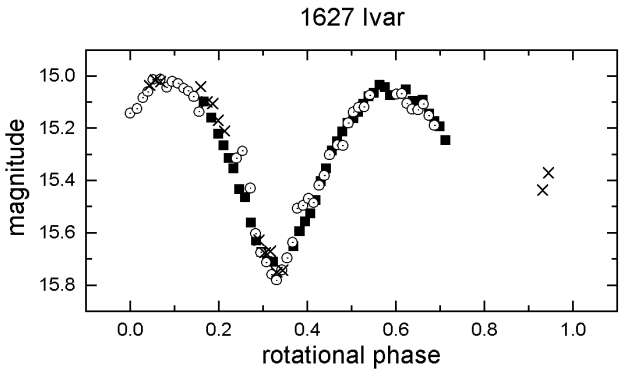,width=\linewidth}
\caption{The composite $R$ lightcurve of 1627 (symbols:
solid squares -- December 14; dotted circles --
December 15; crosses -- December 16).}
\end{center}
\label{1627}
\end{figure}

%Fig. 10.
\begin{figure}
\begin{center}
\leavevmode
\psfig{figure=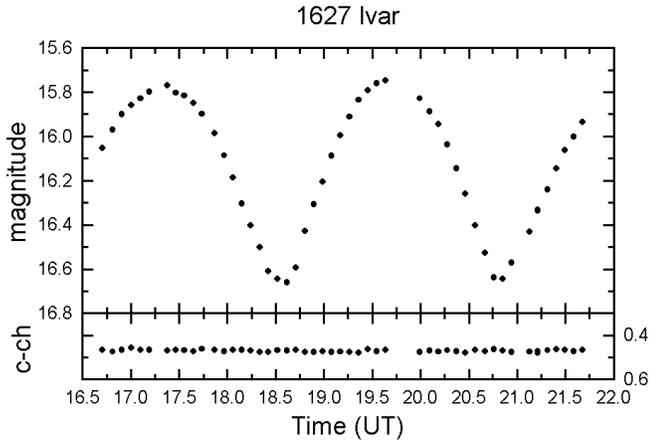,width=\linewidth}
\caption{The $R$ lightcurve of 1627 on January 22, 1999}
\end{center}
\label{1627_99}
\end{figure}

%Table 5
\begin{table}
\begin{center}
\caption{Published photometries of 1627~Ivar}
\begin{tabular} {lrrrlll}
\hline
Date & $\lambda$ & $\beta$ & $\alpha$ & A & t$_{min}$ & ref. \\
\hline
1985 06 13      &  317   &   29    &    48   & 0\fm35 & 46226.750 & (1)\\
1985 08 31      &  15    &   $-$21   &    32   & 0.55 & 46258.703 & (1)\\
1985 10 16      &  4     &   $-$23   &    20   & 0.63 & 46287.184 & (1) \\
1989 05 01-23    & 203   &   25     &   20    & 1.0  & 47647.402 & (2) \\
1989 06 15-23  &  201   &   21    &    51    & 1.12 & 47647.402 & (2) \\
1989 07 14-19  &  213   &   14    &    60    & 1.45 & 47721.565 & (2) \\
1990 05 11-14   &  204   &   25    &    24   & 1.08 & 48029.439 & (3,4) \\
1998 12 14,16   &  76    &   $-$12   &    79   & 0.77 & 51162.295 & p.p. \\
1999 01 26      &  87    &   $-$13   &    18   & 0.92 & 51201.171 & p.p.\\
\hline
\end{tabular}
\end{center}
References: (1) -- Hahn et al. 1989; (2) -- Chernova et al. 1995;
(3) -- Velichko et al. 1990;
(4) Hoffmann \& Geyer 1990
\end{table}

A new amplitude model has been determined after collecting all available
data (Table 5).
The observed amplitudes were reduced to zero solar phase. First of
all, the $m$ parameter was derived from our measurements.
The observed amplitudes in December, 1998 and in January, 1998 were
compared. As the longitudes differ by only 10$^\circ$, and the
difference between the corresponding phases is quite high (13$^\circ$),
the amplitude change can be mostly associated with the phase change.
The result is $m=0.018$.
We have also corrected other amplitudes to zero solar phase and
fitted the amplitude variations along the longitude. The corresponding
parameters are: a/b=2.0$\pm$0.1, b/c=1.09$\pm$0.05,
$\lambda_p=145/325\pm8^\circ$,
$\beta_p=34\pm6^\circ$.
The reduced amplitudes with the determined fit is presented in Fig. 11.
The reliability of this model was
tested by a direct comparison with radar images of Ostro et al. (1990).
This is shown in Fig.\ 12, where we used Fig. 5 taken from
Ostro et al. (1990) with kind permission of the first author.
The similarity is evident.

The O$-$C' method was used to determine the sidereal
period and the sense of the rotation. The results are
P$_{sid}$=0\fd1999154$\pm$0\fd0000003, retrograde rotation with
$\lambda_p=143\pm8^\circ$, $\beta_p=-37\pm6^\circ$ pole coordinates.
The agreement between the poles obtained by different methods is
very good. The sidereal period agrees well with results of
Lupishko et al. (1986) -- 0\fd19991, prograde --, but the senses are
in contradiction. The fitted O$-$C' diagram is presented in Fig.\ 13.

%Fig. 11.
\begin{figure}
\begin{center}
\leavevmode
\psfig{figure=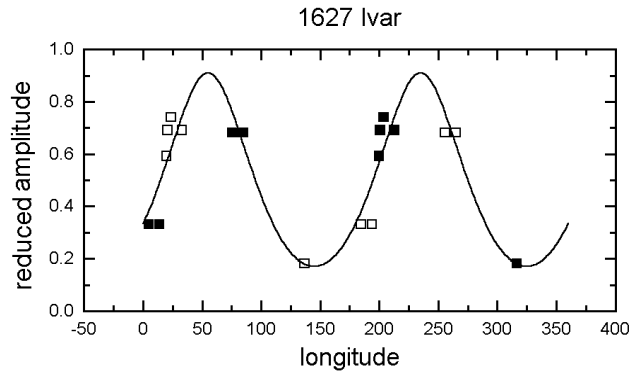,width=\linewidth}
\caption{The reduced amplitudes vs. longitudes with the determined fit
for 1627 Ivar}
\end{center}
\label{1627fit}
\end{figure}

%Fig. 12.
\begin{figure}
\begin{center}
\leavevmode
\psfig{figure=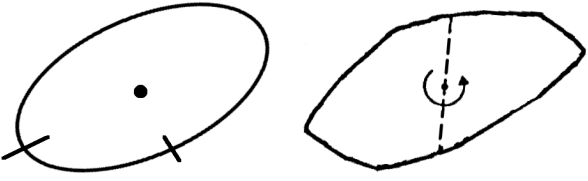,width=\linewidth}
\caption{A pole-on view of the photometric model ({\it left}) and
radar profile ({\it right}) of 1627~Ivar. The small ticks correspond to
the uncertainties of the fit.}
\end{center}
\label{1627mod}
\end{figure}

%Fig. 13.
\begin{figure}
\begin{center}
\leavevmode
\psfig{figure=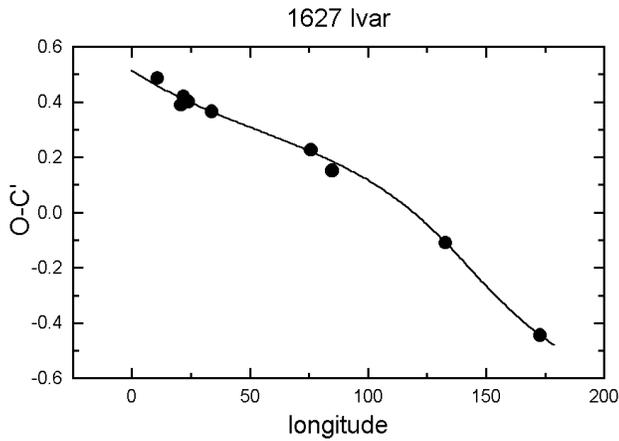,width=\linewidth}
\caption{The observed O$-$C' values fitted with the model for 1627 Ivar}
\end{center}
\label{1627ocfit}
\end{figure}

\bigskip

\bigskip

\bigskip

\bigskip

\noindent {\it 1998~PG}

\noindent The Near Earth Object (NEO) 1998~PG was discovered by the
LONEOS project in Flagstaff, on August 3, 1998. We observed about
80 days after the discovery, in October, 1998.
We found complex, strongly scattering lightcurves (two of them
are shown in Figs.\ 14--15),
which did not show any usual regularity.
Therefore, we performed a conventional
frequency analysis by calculating Discrete Fourier Transform (DFT)
of the whole dataset (Fig.\ 16). Data obtained on October 27 are too noisy,
thus we excluded them from the period determination.

%Fig. 14.
\begin{figure}
\begin{center}
\leavevmode
\psfig{figure=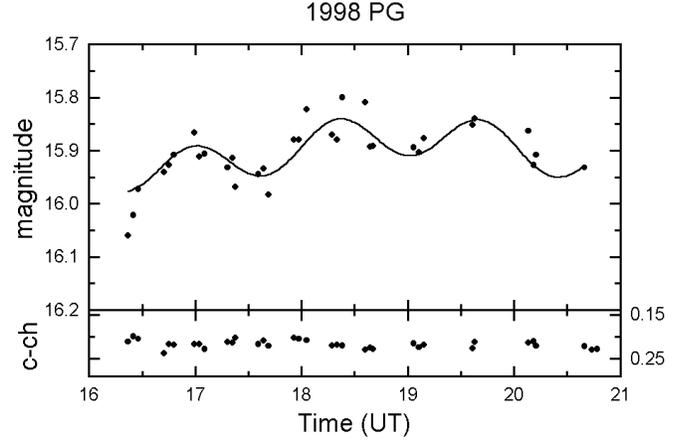,width=\linewidth}
\caption{The observed and fitted $R$ lightcurves of 1998~PG
on October 23, 1998}
\end{center}
\label{pg_23}
\end{figure}

%Fig. 15.
\begin{figure}
\begin{center}
\leavevmode
\psfig{figure=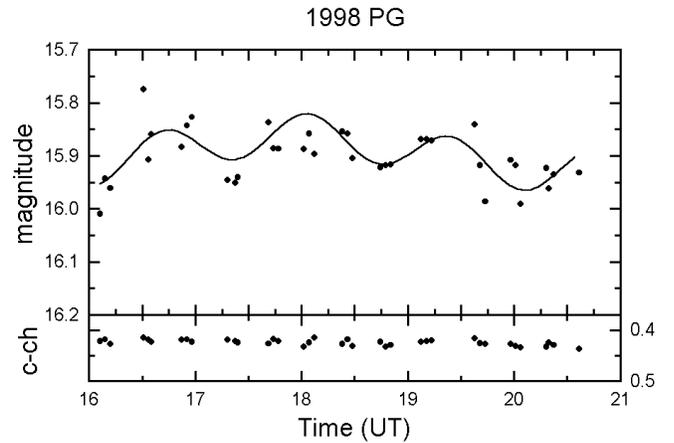,width=\linewidth}
\caption{The same as in Fig.\ 14 on October 26, 1998}
\end{center}
\label{pg_26}
\end{figure}

%Fig. 16.
\begin{figure}
\begin{center}
\leavevmode
\psfig{figure=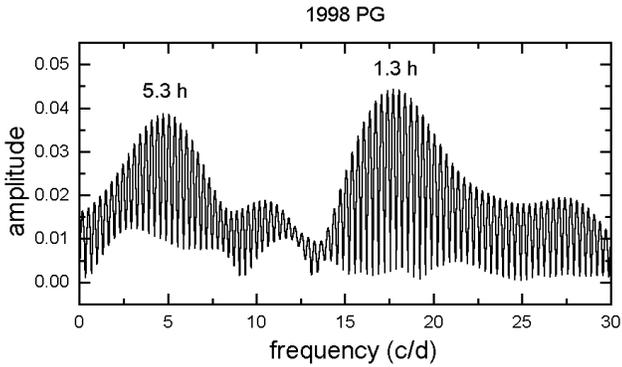,width=\linewidth}
\caption{Fourier spectrum of 1998~PG.}
\end{center}
\label{pg_fou}
\end{figure}

The determined periods are 1\fh3 and 5\fh3, although these values
have large uncertainties (about 10--15\%). Assuming that the
shorter period is due to rotation, we get a rotational period
of 2\fh6. We note that our period values do not contradict
those obtained by P. Pravec and his collaborators, who
found $P_{rot}$=2\fh517 and $P_2\approx$ 7\fh0 (Pravec 1998, personal
communication). The reason for doubly periodic lighcurve can
be precession and/or binarity. The observed rate
of multiperiodic lightcurves among NEOs is quite high (see,
e.g., Pravec 1999), but
the underlying physical processes can only be identified with
more detailed observations than we have on 1998~PG. Therefore,
we conclude that we may have found evidence for precession in 1998~PG, but
other explanations cannot be excluded.

\bigskip

We summarize the resulting sinodic periods, amplitudes and models
in Table 6.

%Table 6
\begin{table*}
\begin{center}
\caption{The determined periods, amplitudes, spin vectors and shapes.}
\begin{tabular} {rllllllll}
\hline
Asteroid      & P$_{sin} (h)$ & P$_{sid} (d)$ & A (mag) & $\lambda_p$ &
$\beta_p$ & a/b & b/c & method \\
\hline
683     & 4.6     & & 0.13       & 15/195$\pm$25 & 52$\pm$15 & 1.15$\pm$0.05 & 1.05$\pm$0.05 & A\\
        & & 0\fd1964156 R & & & & & & O$-$C \\
725     & $\geq$3 & & $\geq$0.4  & -- & -- & -- & -- & A\\
852     & 4.62    & & 0.32, 0.27 & 30/210$\pm$20 & 30$\pm$10 & 2.3$\pm$0.3 & 1.2$\pm$0.2 & A\\
1627    & 4.80    & & 0.77, 0.92 & 145/325$\pm$8 & 34$\pm$6 & 2.0$\pm$0.1 & 1.09$\pm$0.05 & A\\
        & & 0\fd1999154 R & & 143 & $-$37 & & & O$-$C \\
1998~PG & 2.6     & &  0.09      &           &          &            &   & Fourier \\
  ---   & 5.3     & &  0.08      &           &          &            &   & Fourier  \\
\hline
\end{tabular}
\end{center}
\end{table*}

\begin{acknowledgements}
This research was supported by the Szeged Observatory Foundation.
The warm hospitality of the staff of Konkoly Observatory and
their provision of telescope time is gratefully acknowledged.
The authors also acknowledge suggestions and careful reading of the
manuscript by K. West.
The NASA ADS Abstract Service was used to access references.
\end{acknowledgements}

\end{document}